\documentclass[preprint,prb,superscriptaddress,aps]{revtex4}
\usepackage{graphicx}
\usepackage{dcolumn}

\usepackage{amsmath,amssymb}
\usepackage{boxedminipage}
\usepackage{array}

\usepackage{flafter}
\usepackage[mathscr]{euscript}
\usepackage[tight,TABTOPCAP]{subfigure}
\usepackage{verbatim}
\newcolumntype{.}{D{.}{.}{1}}
\begin{document}
%%%%%%%%%%%%%%%%%%%%%%%%%%%%%%%%%%%%%%%%%%%%%%%%%%%%%%%%%%%%%%%%%%%%%%%%%%%%%%%%%%%%%
\title{
Hidden long-range correlations in the ion distribution at the graphite / [bmim][NTf$_2$] electrified interface
}
%%%%%%%%%%%%%%%%%%%%%%%%%%%%%%%%%%%%%

\author{Diego Veloza-Diaz}
\affiliation{Institut f{\"u}r Physik, Johannes Gutenberg-Universit{\"a}t Mainz, Staudingerweg 9, 55128 Mainz, Germany}
\author{Robinson Cortes-Huerto$^{*}$}
\affiliation{Max Planck Institute for Polymer Research, Ackermannweg 10, 55128, Mainz, Germany}
\author{Pietro Ballone}
\affiliation{Max Planck Institute for Polymer Research, Ackermannweg 10, 55128, Mainz, Germany}
\author{Nancy C. Forero-Martinez$^{*}$}
\affiliation{Institut f{\"u}r Physik, Johannes Gutenberg-Universit{\"a}t Mainz, Staudingerweg 9, 55128 Mainz, Germany}

\begin{abstract}
A capacitor consisting of the [bmim][NTf$_2$] ionic liquid (IL) confined in between planar graphite electrodes has been investigated by molecular dynamics based on an all-atom, unpolarizable force field. Despite a few peculiarities due to the size  and complexity of the ions, properties such as the density of ions throughout the capacitor, the screening of the surface charge on the electrodes by the IL and exact sum rules for the radial distribution functions of cations and anions generally comply with the results of time honored theories of the electrostatic double layer. This soothing regularity may conceal hidden correlations still compatible with the static screening rules, propagating far inside the IL the information on the state of charge of the capacitor. Evidence in this respect might have been detected by vibrational spectroscopy (see, for instance, {\it Langmuir} {\bf 2021}, {\it 37}, 5193-5201) showing changes in optical properties of the IL far from the charged electrodes. We show that grouping the [bmim]$^+$ and [NTf$_2$]$^-$ ions into instantaneous neutral pairs reveals an intriguing long range ordering of ions normal to the interface, driven by the capacitor state of charge. These correlations manifest themselves through the parallel orientation of the dipole moments of the neutral ion pairs. We speculate that this effect changes the average intensity of fluctuating electric fields deep in the IL, while average, static fields vanish in agreement with well established screening laws. This effect, which could change the spectroscopic properties of the IL, is present in the simulated [bmim][NTf$_2$] / graphite capacitor, but too small to be unambiguously confirmed by the present simulations with a safe margin over the error bar. The conceptual interest in these effects, however, will motivate further studies of the same or similar electrode / ionic liquid interfaces.
\vskip 1.0truecm
\noindent
{\bf $^{ *}$ Corresponding authors: corteshu@mpip-mainz.mpg.de, nforerom@uni-mainz.de}
\end{abstract}

\maketitle

\section{Introduction}
\label{intro}
The electrostatic double layer is a unifying  concept that plays a fundamental role in the chemical-physics of interfaces.
As such, it also represents a crucial aspect of most electro-chemical devices and related processes.\cite{bock} In the last two
decades the interest in this time honored topic has been enhanced by two developments, i.e., the expanded role of 
electro-chemical energy conversion and storage applications (batteries, fuel cells, super-capacitors) as well as the surge of
activity concerning ionic liquids (ILs) which offer tantalizing opportunities for increasing efficiency, durability, safety
and environmental compatibility of electrochemical processes and devices. References on those two research fields are too 
numerous to cite, and we point to the extensive discussions in Ref.~\onlinecite{korn1, korn2, korn3}.

\begin{figure}[!htb]
%\begin{minipage}[c]{\textwidth}
%\vskip 0.7truecm
%\begin{center}
\includegraphics[scale=1.00,angle=-0]{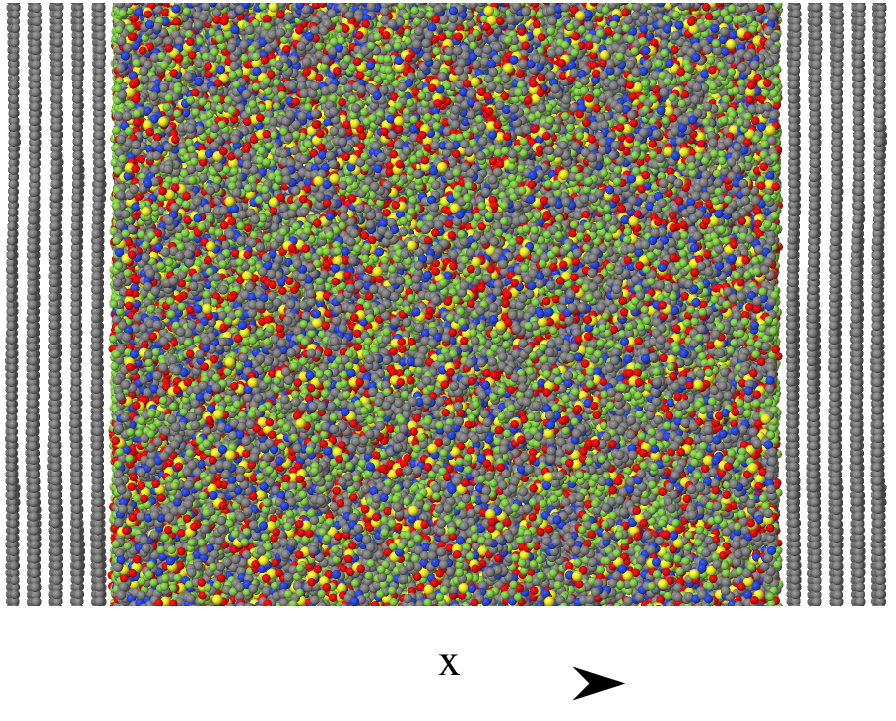}
%\vskip 1.0truecm
\caption{The dry graphite/[bmim][NTf$_2$]/graphite capacitor. Gray, blue, red, green and yellow dots represent C, N, O, F and 
S atoms, respectively. Hydrogen atoms not shown. The regular vertical layers of C atoms are part of the graphite electrodes. The
right-most graphite layer in the figure is the periodic replica of the left-most layer.
}
\label{capac}
%\end{center}
%\end{minipage}
\end{figure}

The long range of the Coulomb charge-charge interactions poses strict conditions on the structure and dynamics of the
electrostatic double layer that forms at the interface between an electrolyte and an electrode. In particular, the width 
of the spatial charge separation at the interface is limited by screening to a few characteristic lengths, i.e., multiples of a
screening length $\lambda$, whose value depends primarily on the thermodynamic state of the electrolyte. Beyond this (usually 
microscopic) length, the electrolyte appears to be largely {\it bulk-like}.

\begin{figure}[!htb]
%\begin{minipage}[c]{\textwidth}
%\vskip 0.7truecm
%\begin{center}
\includegraphics[scale=0.55,angle=-0]{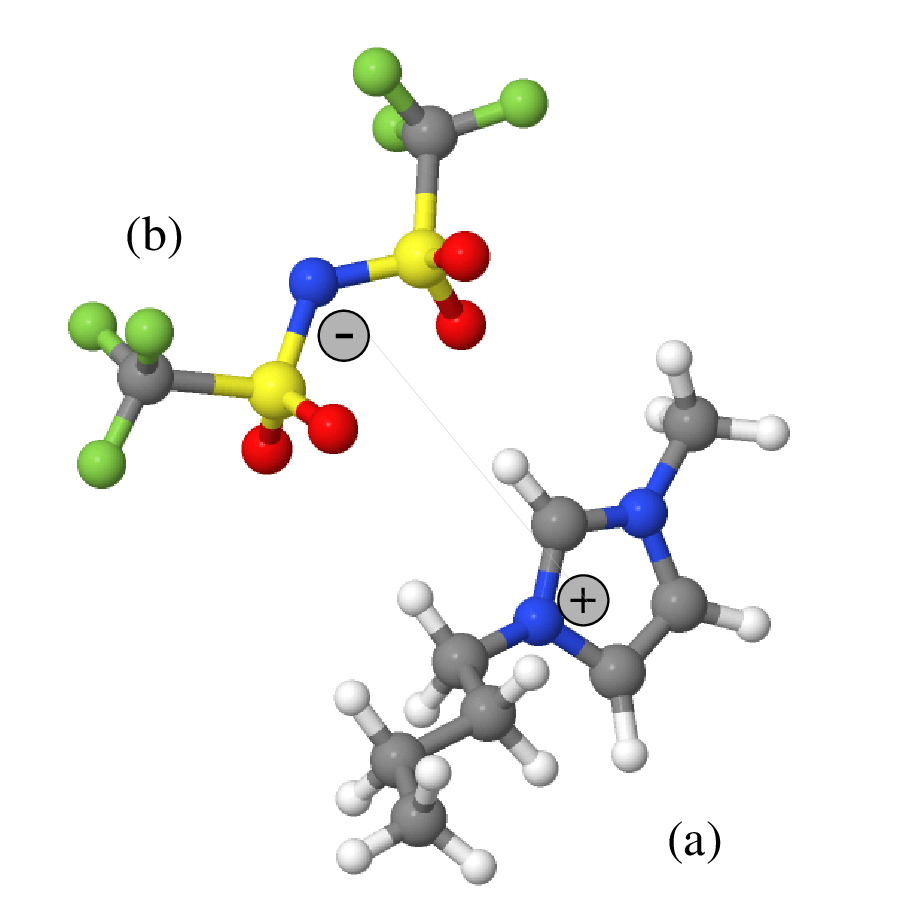}
%\vskip 1.0truecm
\caption{Atomistic structure of the: (a) [bmim]$^+$ and: (b) [NTf$_2$]$^-$ ions. The color coding is the same as in 
Fig.~\ref{capac} The picture shows a generic configuration from a MD trajectory at $T=300$ K. The gray circle with the plus and
minus sign are located at the position of the coarse grained representation of the ions obtained by applying Eq.~\ref{cg}.
}
\label{bmimntf2}
%\end{center}
%\end{minipage}
\end{figure}

Given this generally accepted picture, it came as a surprise the revelation by optical measurements of characteristic changes in
the Raman response of imidazolium-based ionic liquid enclosed between the plates of a static capacitor, driven by the applied 
electric field and affecting the spectroscopic properties of the IL electrolytes far from the 
interfaces.\cite{CARS, expLangmuir} To our knowledge, up to now, effects of this type have been revealed on electrolyte samples
made of:\cite{expLangmuir} 1-butyl-3-methyl-imidazolium bis(trifluoromethylsulfonyl)imide ([bmim][NTf$_2$]); 
1-butyl-3-methyl-imidazolium hexafluorophosphate ([bmim][PF$_6$]); 1-ethyl-3-methyl-imidazolium 
bis(trifluoromethylsulfonyl)imide ([emim][NTf$_2$]) as well as 1-alkyl-3-methylimidazolium hexafluorophosphates 
([C$_n$mim][NTf$_2$], $n=4$, $6$, $8$).\cite{CARS}

Since the strict confinement of the interfacial charge perturbation is supported by very robust arguments and is verified by a 
large number of experimental, theoretical and computational studies, we look for the responsible of this long range ordering at 
the second simpler and most likely suspect, represented by the combination of ions into neutral dipolar aggregates, which might 
display long range correlation effects. As the testbed for this analysis we have chosen the interface between a graphite 
electrode and a paradigmatic IL, consisting of the 1-butyl-3-methylimidazolium cation ([bmim]$^+$) and 
bis(trifluoromethylsulfonyl)-imide anion 
([NTf$_2$]$^-$), i.e., [bmim][NTf$_2$]. This compound is mildly hydrophobic but retains a not-completely-negligible water 
concentration  of up to about $25$ molar \%, which however corresponds to only less that $1.5$ wgt\% due to the mass disparity 
of water and [bmim][NTf$_2$]. Samples of dry and water-contaminated [bmim][NTf$_2$] samples with and without confining graphite
electrodes have been simulated by molecular dynamics (MD) using an atomistic, semi-empirical force field model. We chose 
[bmim][NTf$_2$] for our investigation, despite its complexity, because it has received a considerable  interest as a solvent
electrolyte\cite{electrochem} and because relatively abundant experimental and computational literature is available on its 
physical-chemistry properties. Moreover, the same effect might not be observable in simpler (either realistic or idealised) 
models because size and shape of the electrolyte ions might play a role in the propagation of correlation effects far from the 
interface. On the other hand, graphite has been selected because it is indeed used to make electrodes, and especially because
its interaction with organic species are covered by the same type of interatomic force fields used for the ions themselves,
unlike metal electrodes, whose typical empirical potential energy models are not easy to combines with those of ILs.

The main contribution of the present study is the development and calibration of a novel approach to analyse the MD 
trajectories, which, for each configuration, defines a population of {\it instantaneous} dipoles made from IL ions in the least
arbitrary way. Achieving this goal will allow us to quantify correlations among dipoles as a function of the electron density 
on the electrode, of temperature and of concentration of water contamination, whose presence is due to the hygroscopic character
of [bmim][NTf$_2$]. The MD of the interface, that represents the required background of this investigation, provides a wealth of
additional information on the electrostatic properties of the interface, including an estimation of the differential capacitance
of the interface, and its dependence on temperature, pressure, surface charge density on the electrode, and concentration of 
contaminant water. Since the discussion of all these aspects could obscure the major novelty of the present study, we defer the 
full presentation of our investigation to a later and more comprehensive paper, devoting the present document to a concise 
report of the main result, represented by the presence of dipole correlations far from the interface, where the average charge 
density and electric field vanish. 

\section{Model and Method}
\label{method}
The subject of the major portion of the simulations is a planar capacitor consisting of two parallel graphite plates orthogonal 
to the $\hat{x}$ axis, delimiting a relatively thick ($\sim 11$ nm) [bmim][NTf$_2$] layer, see Fig.~\ref{capac}. The thickness
of the [bmim][NTf$_2$] slab is such to overlap, at least to a limited extent, with the range ($10-100$ nm) of the long range 
correlations found in similar experimental systems. Both graphite and the IL are described at the atomistic level. The potential
energy surface of the system is represented through an atomistic, empirical force field of the Amber form,\cite{amber} 
parametrised at first using the Gromos model in its 54a7\cite{54a7} version for the [bmim][NTf$_2$] component, \cite{gromos} and
the SPC model for water.\cite{spc} In practice, the topology files for the [bmim]$^+$ and [NTf$_2$]$^-$ ions, whose structure is
shown in Fig.~\ref{bmimntf2}, have been obtained 
using the automated topology building utility made available by the ATB website.\cite{atb} At the beginning, all the charges on 
the graphite carbon atoms are set to zero, to represent uncharged electrodes. In the case of electrodes of non-vanishing surface
density $\sigma$, the atomic-like charges of graphite are distributed in contributions $\pm \delta q$ of equal size and 
opposite sign on the outermost layers of the graphite block, as detailed in the following Sec.~\ref{interf}.

The force field for graphite has been built by generating the topology files for coronene (C$_{24}$H$_{12}$) using again Gromos 
54a7 and ATB, then copying the stretching, bending, dihedral and improper dihedral parameters for aromatic carbon to the 
graphite topology file. The 2D graphite periodicity along the plane of the electrode is built into the topology file by listing
all the bonded C pairs including those that connect through pbc.

\begin{figure*}[!htb]
\begin{minipage}[c]{\textwidth}
\vskip 0.7truecm
\begin{center}
\includegraphics[scale=0.35,angle=-0]{graph.eps}
\vskip 1.0truecm
\caption{(a) Honeycomb structure of the single graphite layer (graphene).
(b) stacking of A (red) and B (black) layers whose repetition is makes the finite graphite slab representing the planar plates 
of the capacitor. Slight irregularities in the atomic positions are due to the fact that the configuration has been taken from
an MD simulation at $T=300$ K.
}
\label{graphite}
\end{center}
\end{minipage}
\end{figure*}

Test simulations of homogeneous (3D periodic boundary conditions, pbc) [bmim][NTf$_2$] systems have shown that the 
self-diffusion constant of both [bmim]$^+$ and [NTf$_2$]$^-$ as computed by the Einstein relation are severely underestimated. 
This is a know problem of unpolarisable IL force fields,\cite{diffu} such as Gromos 54a7, and, in the present case 
the problem might be exacerbated by a non-negligible over-estimation of the homogeneous IL density. The diffusivity problem 
sometimes is solved by scaling all atomic charges by a $0.8-0.9$ factor,\cite{08, 09} justified in terms of atomic 
polarizability or partial charge 
transfer among cations and anions (which can also be seen as a form of electronic polarizability). These effects, however, are
likely to be different in the bulk and at an interface. Moreover, any ion that (hypothetically) leaves the condensed phase needs 
to have an integer charge. Lacking a suitable model to include the charge dependence on the instantaneous configuration and on 
the thermodynamic state in a consistent and computationally convenient form, it has been decided to keep unchanged the formal 
charge $+1 e$, $-1 e$ of cations and anions, respectively. This decision implies accepting a general slowing down of the ion 
dynamics with respect to the experimental system. The density over-estimation, instead, has been addressed by an empirical 
tuning of the Lennard-Jones (LJ) pair potential, hoping to improve at the same time the diffusivity aspect. To this aim, all 
Lennard-Jones radii $\sigma_{ij}$ of [bmim][NTf$_2$] atoms have been expanded by a constant factor of $1.03$ (+3\%), recovering 
the experimental density of pure and homogeneous [bmim][NTf$_2$] at $T=300$ K. The homogeneous liquid properties of 
[bmim][NTf$_2$] estimated by this model are reported and briefly discussed in Sec.~\ref{results}. The aromatic carbon 
description of graphite has been left unchanged at the Gromos 54a7 level, since it has little effect on the IL density and 
dynamics even in its proximity besides confining the IL to the inside of the model capacitor.

Water is present in a few of the samples at relatively low concentration (20 molar \%, corresponding to less that $1$ wgt\%).
At these concentrations, we verified that water molecules move in the [bmim][NTf$_2$] liquid virtually independently from each 
other, their clustering being limited to a few water dimers and very rare trimers over the entire sample. Hence, their evolution
is determined primarily by the structure and dynamics of the IL fraction, whose properties, in turn, are affected by the water 
contamination by a relative variation that exceeds the mass fraction concentration of water in the system. In the present 
computation water is described by the three-sites FBA/$\epsilon$ model of Ref.~\onlinecite{FBAoe}, which represents water 
molecules as flexible, with bonded and non-bonded parts of the force field only slightly re-parametrised with respect to the 
rigid (SPC\cite{spc}) and flexible (SPC/Fw\cite{fw}) three-sites simple charge water model. The FBA/$\epsilon$ model has been
selected because already used in combination with a [bmim][NTf$_2$] model of the same overall density.\cite{FBAbmim}

All planar capacitor samples are enclosed in an orthorhombic box, with periodic boundary conditions applied. More precisely, at
first samples are periodically replicated in 2D along the $(yz)$ plane. Then, despite the inherent 2D character of the 
capacitor, the electrostatic interactions have been computed using 3D Ewald sums upon replicating the capacitor also along the 
$\hat{x}$ direction, because 2D Ewald sum routines are generally considered to be less efficient than their 3D counterparts. No 
explicit test, however, has been carried out to quantify or even verify this statement for the present case. the periodicity
along $x$ is equal to the combined width of the graphite (9-layer) block and of the [bmim][NTf$_2$] slab. In other words, no gap
is left between the sample and its nearest replicas along $x$. The thickness of the graphite slab ($\sim 30$ \AA\ ) is the major
aspect that limits the spurious interactions of the charged interfaces and [bmim][NTf$_2$] slab across the pbc boundary.

The artificial periodicity along $x$, however, has a far more significant effect, that needs to be accounted for. An isolated,
single planar capacitor of charge $Q=\sigma A$, where $\sigma$ is the surface charge density on the positive electrode and $A$ 
is the electrode area, is globally neutral. Hence, the electric field on its outside vanishes, and symmetry considerations show 
that inside the capacitor the electric field is oriented normal to the plates. Then, application of Gauss' theorem immediately 
gives that the electric field strength on its inside is a constant equal to $4\pi \sigma$. For an infinitely periodic sequence 
of capacitors, however, the analysis is not so simple, because there is no simple way to locate a surface of vanishing electric 
field (${\bf E}=0$) from which to start applying Gauss' theorem to compute ${\bf E}(x)$ anywhere else in space. Seen in a 
different but equivalent way, there is no reason why the electric field within the graphite blocks separating neighbouring 
replicas has to vanish. Hence, the electric field inside the liquid electrolyte, although still constant and 
directed along $x$, is not equal to $4\pi \sigma$.\cite{chris} The precise value of ${\bf E}(x)$ can be computed analytically 
based on the Ewald sum formalism, but it depends on boundary condition parameters set in the simulation code, whose value in not
easily determined. To avoid this ambiguity, the electric field ${\bf E_0}=E_0(x)\hat{x}$ along the $x$ coordinate within the 
capacitor and within the graphite slab has been computed using Gromacs considering an empty capacitor whose charged plates 
(represented by the charge graphite layers) are located at the distance used in the actual simulations with the ionic liquid. 
This {\it empty capacitor} curve $E_0(x)$ has been used throughout the paper to compute the electrode contribution to the
position-dependent electric field ${\bf E}(x)$ and electrostatic potential ${\bf \phi}(x)$ through the capacitor.

Most simulations have been carried out in the NPT ensemble, using the Parrinello-Rahman approach. The remaining few simulations 
have been done in the NVT ensemble, mainly to have a fixed reference frame to compute and plot density distribution functions 
perpendicular to the interface. The NVT ensemble has also been used to simulate free standing dry and wet ionic liquid slabs, 
exposing two parallel (on average) IL/vacuum interfaces.

The graphite plates, in particular, consist of hexagonal layers of carbon, with aromatic sp$^2$ bonds among them. A solid-like
graphite slab is formed considering a stack of 9 layers in the ABABABABA configuration (see Fig.~\ref{graphite}). It has been 
verified that the NPT simulation of the identically neutral 9-layers graphite slab in pbc at 1 bar (no vacuum) and $T=300$ K 
reproduces fairly well the structural parameters of graphite, included the wide layer-layer separation $d\sim 3.5$ \AA\ .

All simulated systems, including the homogeneous liquid [bmim][NTf$_2$] samples, consist of the same number of cations and 
anions. The electrolyte slab, therefore, is globally neutral. Since also the whole sample, which is 3D periodic, needs to be 
neutral, the two interfaces delimiting the electrolyte may only carry zero or opposite surface charge densities $\pm \sigma$. At
first, the charge on each graphite C atom is identically zero.  Then, any non-vanishing charge density on the graphite surfaces 
will be represented by atomic charges $q_{gr}=\pm \delta q$ of equal modulus
and opposite sign on the two surfaces. Since graphite is a conductor, the static $\pm \sigma$ charges on the two sides of the 
graphite, in reality, would not be stable with respect to charge recombination. However, the electrical conductivity 
perpendicular to the graphite planes is about $10^3$ times lower than along the planes. Therefore, the simple charging model is 
justified by neglecting the electrical conductivity across the graphite slab, and assuming that the two outermost C-planes are 
connected to the two poles of a battery.

All simulations have been carried out using the Gromacs\cite{gromacs} simulation package, version 2019.3. The analysis of 
trajectories has been made using home-made programs. These have been used to compute number and charge density profiles along 
the direction normal to the graphite planes, ion-ion radial distribution functions, self-diffusion coefficient based on the 
Einstein equation, orientation correlations among the ions and with respect to the planar IL/graphite interfaces. The average
electrostatic potential along the same normal direction has been computed by integrating the electric field inside he capacitor,
which is the sum of the [bmim][NTf$_2$] contribution determined from the average charge density profile of the ions, and the
contribution from the charge on the electrodes, computed by Gromacs for the empty capacitor, as explained a few paragraphs
above in this same section.

\begin{figure}[!htb]
%\begin{minipage}[c]{\textwidth}
%\vskip 0.7truecm
%\begin{center}
\includegraphics[scale=0.60,angle=-0]{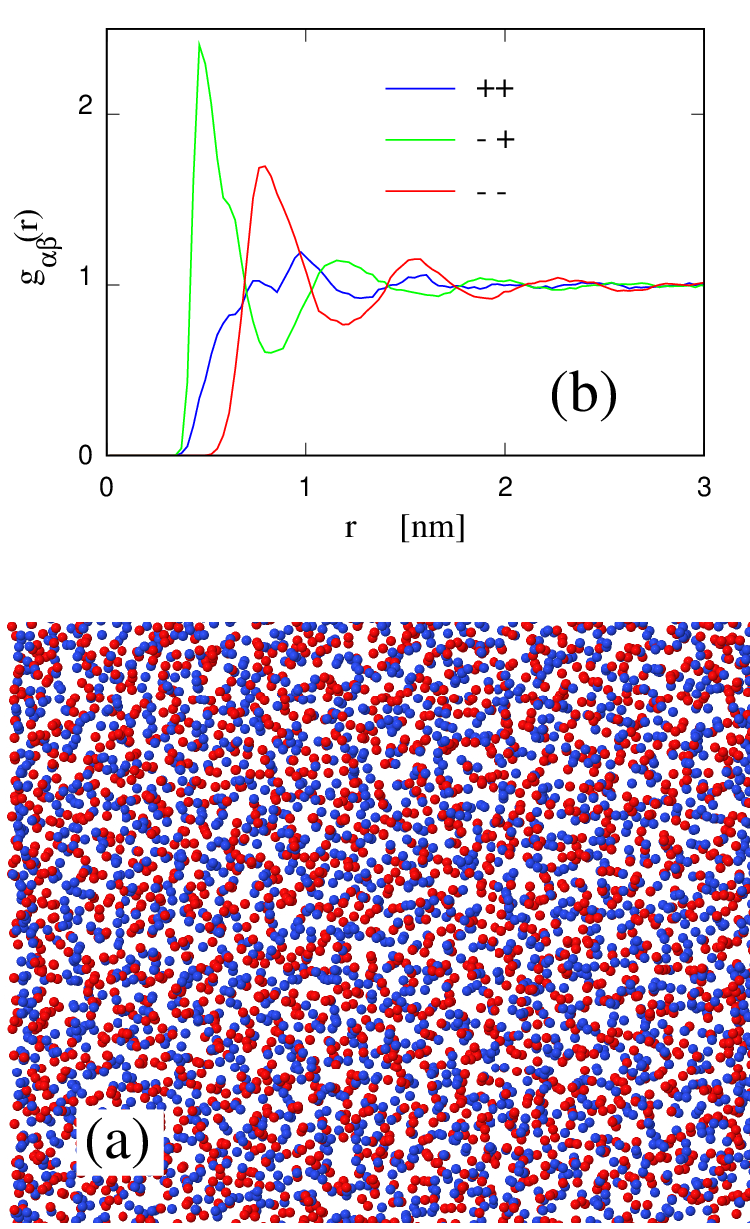}
%\vskip 1.0truecm
\caption{(a) Snapshot of the ionic liquid part of Fig.~\ref{capac} coarse grained through Eq.~\ref{cg}. Red dots: cations; blue
dots: anions. (b) radial distribution functions of cations and anions
at the same conditions of panel (a). Only the ions in the central $5$ nm of the ionic liquid slab have been considered as 
centres to compute the corresponding $g_{\alpha \beta}(r)$.
}
\label{coarse}
%\end{center}
%\end{minipage}
\end{figure}

A crucial aspect of the trajectory analysis is represented by the definition of dipoles within the ionic liquid, and their
localisation in space. The first step in this process consists in identifying the position of each ion with its centre of 
charge:
\begin{equation}
{\bf R^{(Q)}_{\alpha}}=\frac{\sum_{i \in \alpha} \mid q_i \mid {\bf r_i}}{\sum_{i \in \alpha} \mid q_i \mid }
\label{cg}
\end{equation}
whose definition is apparently analogous to that of the centre of mass of a set of particles. The choice of emphasising the 
charges instead of the masses is due to our aim of identifying each ion with its charge. This definition is not as unanimously 
accepted nor extensively used as that of the centre of mass. In this context, however, it provides a simple way to coarse grain 
the charge distribution of [bmim][NTf$_2$], which is fairly complex and difficult to visualise and interpret in its atomistic 
detail. For instance, the configuration of cations and anions corresponding to the snapshot in Fig.~\ref{capac} is shown in 
Fig.~\ref{coarse} (a), highlighting the drastic simplification of the IL structure achieved by the coarse graining expressed by 
Eq.~\ref{cg}. The radial distribution functions (rdf) computed for the binary mixture consisting of cations and anions and shown
in Fig.~\ref{coarse} (b) present all the features expected for a Coulombic fluid, including: the charge alternation around each 
ion chosen as the origin of the rdf; a tall anion-cation $g_{+-}$ peak, and a strong non-additivity in the distance of closest 
approach for the $++$ (long), $+-$ (short) and $--$ (long) ion pairs. The coordination numbers ($\sim 6$ for cation-anion and 
$\sim 12$ for anion-anion, while the cation-cation cannot be distinguished) are vaguely reminiscent of the NaCl structure.

The method to highlight subtle correlations in the distribution of ions in the simulated trajectories is somewhat complex and
consists of a few successive steps. For this reason, the partition of ions into neutral pairs is defined in the following 
Sec.~\ref{results} using actual simulation results to illustrate this protocol.

\section{Results}
\label{results}
\subsection{Properties of the homogeneous samples}
Homogeneous samples of dry and water-contaminated [bmim][NTf$_2$] have been simulated to validate the force field and to 
estimate baseline values for density, diffusion constant of cations and anions, and basic structural properties such as the 
radial distribution functions of cations, anions and water. Each sample consisted of $3072$ neutral ([bmim]$^+$ and 
[NTf$_2$]$^-$, see Fig.~\ref{bmimntf2}) ion pairs, and up to $640$ water molecules in a cube periodically replicated in space.

Since the experimental density of [bmim][NTf$_2$] at $T=300$K was fitted by increasing the LJ radii, the computed average 
density of the model, $\rho=1.4602 \pm 0.0002$ g/cm$^3$, falls into the range of experimental values (from $\rho=1.44$ to $1.47$
g/cm$^3$) measured in experiments at the same temperature. The thermal expansion coefficient is somewhat underestimated,
being $K_T=2.72\ 10^{-4}$ K$^{-1}$ for the model, versus the experimental $K_T=6.4\ 10^{-4}$ K$^{-1}$.
At the $20$ molar \% concentration, water appears to be soluble in the re-parametrised [bmim][NTf$_2$] model, in agreement with
experiments.

\begin{figure}[!htb]
%\begin{minipage}[c]{\textwidth}
%\vskip 0.7truecm
%\begin{center}
\includegraphics[scale=0.70,angle=-0]{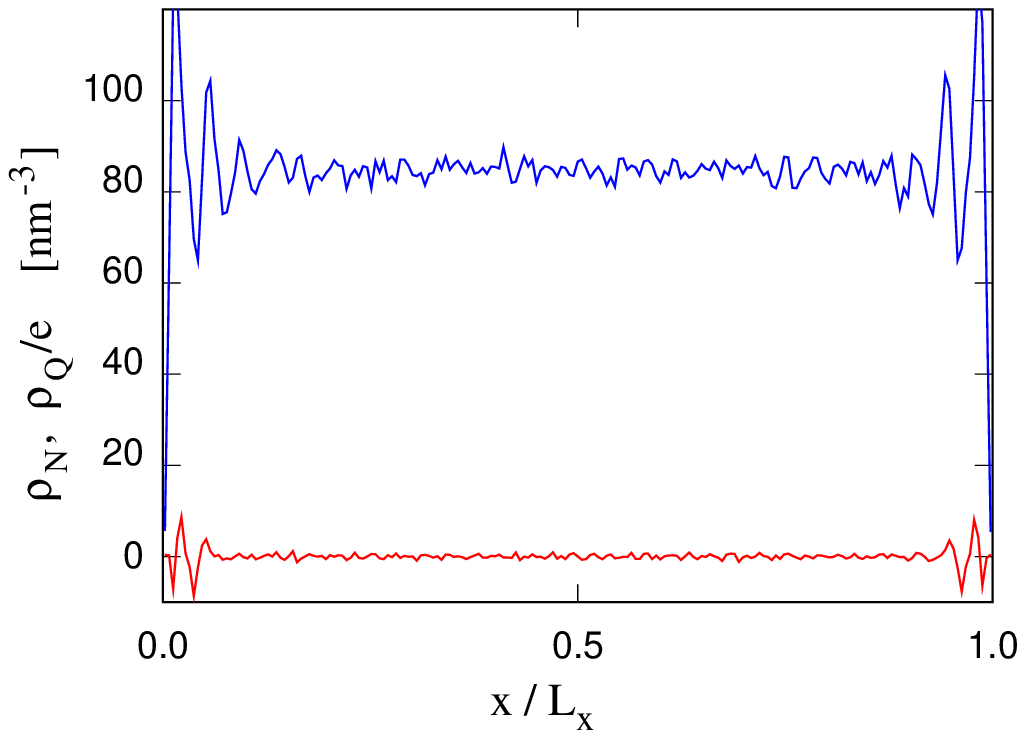}
%\vskip 1.0truecm
\caption{Number ($\rho_N(x)$, blue line)  and charge ($\rho_Q(x)$, red line) density profiles across the uncharged ($\sigma=0$) 
capacitor. $L_x$ is the average separation of the outermost graphite layers measured across the [bmim][NTf$_2$] slab.
}
\label{profnx}
%\end{center}
%\end{minipage}
\end{figure}

The results show that the self-diffusion coefficient of cations and anions in pure, 3D homogeneous [bmim][NTf$_2$] at $T=300$ K,
although higher than that of the original Gromos 54a7 model, is still significantly below the experimental value measured by 
NMR.\cite{perk} Moreover, and possibly more importantly, the estimated value
depends on the time scale of the simulation, revealing a slightly sub-diffusive behaviour. Nevertheless, the mean square 
displacement is largely linear over a $> 50$ ns interval, with an estimated value of $D$ of the order of $10^{-9}$ cm$^{2}$/s 
for both anions and cations to within  a large statistical error of $30$ \%. This computed value has to be compared to
the experimentally measured $10^{-7}$ cm$^{2}$/s, see the diffusion constant of [bmim][NTf$_2$] measured by NMR as 
reported in Ref.~\onlinecite{perk}. We accept this discrepancy since unpolarizable ions are still the only way to simulate 
system sizes in excess of $10^5$ atoms at an acceptable cost. Addition of water at concentration as low as $1$ wgt\% hardly 
affects the mobility of the ions. The self-diffusion coefficient of water in [bmim][NTf$_2$] at 20 wgt\% concentration and
$T=300$K turns out to be $(2.5 \pm 0.5)\ 10^{-7}$ cm$^2$/s, i.e., two orders of magnitude lower than in pure water at the same 
$T$, $P$ conditions.

\begin{figure}[!htb]
%\begin{minipage}[c]{\textwidth}
%\vskip 0.7truecm
%\begin{center}
\includegraphics[scale=0.70,angle=-0]{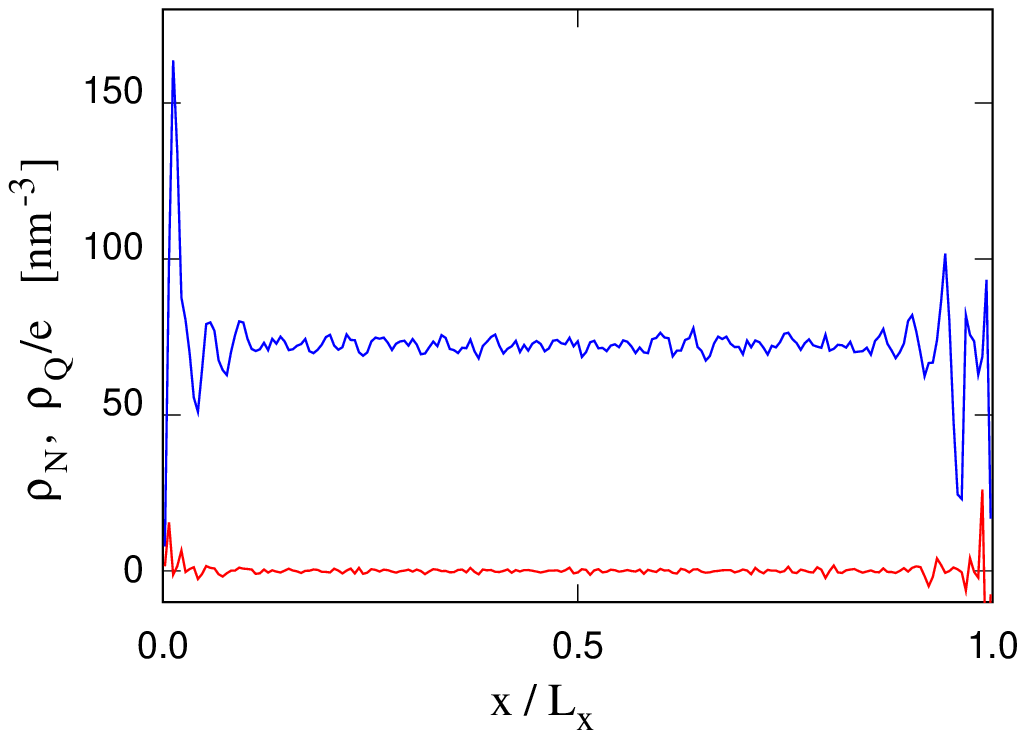}
%\vskip 1.0truecm
\caption{Number ($\rho_N(x)$, blue line) and charge ($\rho_Q(x)$, red line) density profiles across the capacitor with 
$\sigma=2$ e/nm$^2$ surface charge density. 
$L_x$ is the average separation of the outermost graphite layers measured across the [bmim][NTf$_2$] slab.
Notice the change of scale with respect to Fig.~\ref{profnx}.
}
\label{profqx}
%\end{center}
%\end{minipage}
\end{figure}

\subsection{Structural, dynamic and electrostatic properties of the graphite / [bmim][NTf$_2$] interface} 
\label{interf}

The primary focus of the simulations is a planar capacitor consisting of $3072$ ([bmim]$^+$, [NTf$_2$]$^-$) neutral ion pairs,
enclosed in between two electrodes that are represented by the opposite faces of a  graphite slab, whose hexagonal planes are 
parallel to the interface. This basic sample is illustrated in Fig.~\ref{capac}. Each simulation cell is orthorhombic, and
excluding the graphite slab, each IL sample is nearly cubic.

The reference frame has been oriented in such a way that the interface is orthogonal to the $x$ axis. For this reason, local 
properties such as density and electrostatic potential  profiles are plotted as a function of $x$.  
To simulate charged capacitors, small charges of up to a few $10^{-2}$ e (where e is the H$^+$ charge) have been equally 
assigned to all the atoms of the two most external layers of graphite, qualitatively reflecting the picture provided by
{\it ab-initio} (density functional theory) computations, see Ref.~\onlinecite{charging}. The charges are of equal magnitude and
opposite sign on the two charged layers. Since also the IL slab is globally neutral, the whole sample is neutral.
Since, according to the simulation results, the 2D number density of C atoms in the graphite layers at $T=300$ K is
$37.08$ nm$^{-2}$, a charge density of 0.05 e per graphite layer corresponds to a sizeable surface charge density of nearly
$2$ e/nm$^{-2}$.

\begin{figure}[!htb]
%\begin{minipage}[c]{\textwidth}
%\vskip 0.7truecm
%\begin{center}
\includegraphics[scale=0.70,angle=-0]{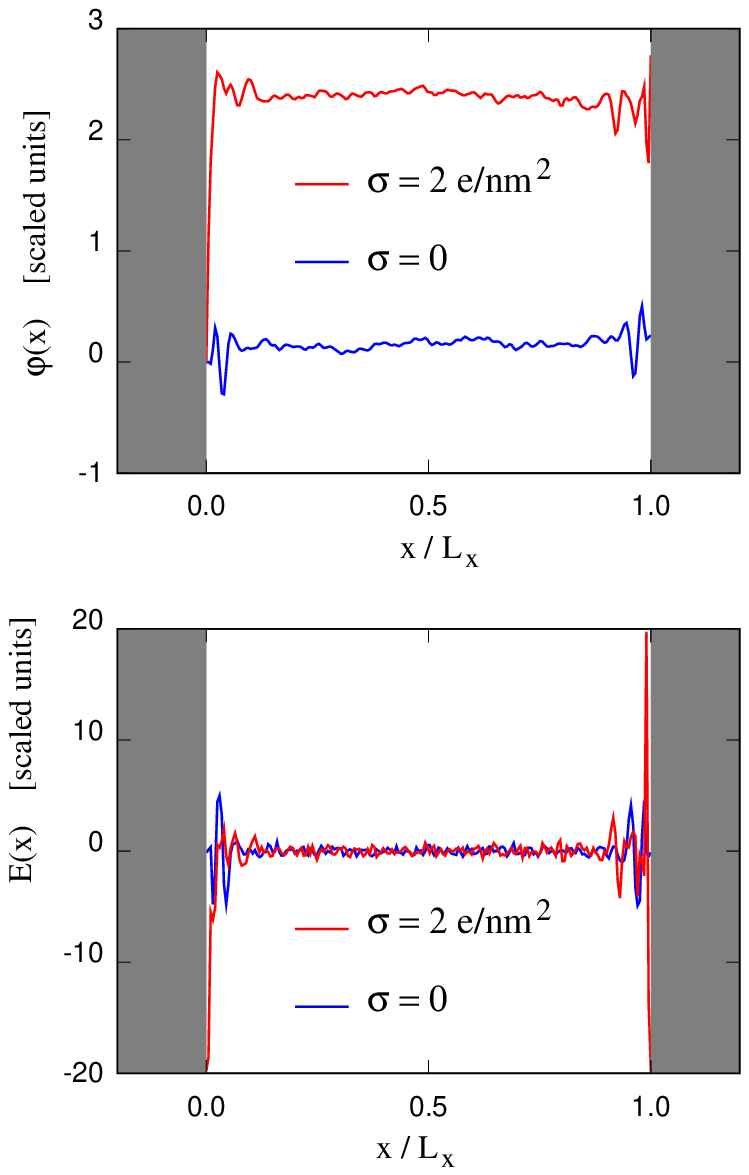}
%\vskip 1.0truecm
\caption{Electric field and electrostatic potential  as a function of the $x$ coordinate across the capacitor for
two different values of the surface charge $\sigma$. The gray rectangles represent the graphite electrodes.
$L_x$ is the average separation of the outermost graphite layers measured across the [bmim][NTf$_2$] slab.
The scaled units for $E(x)$ and $\varphi (x)$ are explained in the text.
}
\label{elex}
%\end{center}
%\end{minipage}
\end{figure}
The simukation results for the number and charge density profile as a function of $x$ across the capacitor are shown in
Fig.~\ref{profnx} and Fig.~\ref{profqx}. The charge profile across the capacitor, in particular, is the required ingredient to 
compute the electrostatic potential as a function of $x$, displayed in Fig.~\ref{elex} for the $T=300$ K sample at $\sigma=0$ 
and at $\sigma=2$ e/nm$^2$. As expected,
the electrostatic potentials is flat and the electric field vanishes over a wide central region of the capacitor, independently 
from the surface charge.

%The output of the 
%electrostatic potential analysis will provide data to compute the differential capacitance per unit area of the interface, 
%defined as:
%\begin{equation}
%C_d=\left( \frac{\partial \sigma }{\partial \varphi}\right)_{T,P}
%\end{equation}
%that represent a crucial parameter characterizing the electrostatic double layer at the interface, measuring the width of
%the double layer itself.

\subsection{Subtle electrostatic correlations deep into the bulk-like interior of the electrolyte}

The results of the previous subsections show that the {\it average} electric field vanishes away from the interfaces. The 
presence of electro-optical effects sufficiently strong to be detected by spectroscopy, however, does not require a 
non-vanishing static electric field. A detectable Stark effect, for instance, can result from a change in the average square 
modulus $E^2$ of the fluctuating electric field acting on each ion. Such a square modulus does not vanish even in bulk 
[bmim][NTf$_2$], and changes in its value can result from subtle correlations compatible with the strict screening sum rules 
that exclude net charge separation and non-vanishing average electric field away from the interface. Thus, excluding the 
violation of these rules, the present investigation focuses on possible correlations among electrostatic dipoles.

Such an analysis can follow a variety of leads, but two aspects need to be retained. First, one needs to define dipoles inside 
an ionic fluid, which is certainly a challenging undertaking since the [bmim]$^+$ and [NTf$_2$]$^-$ ions are mobile. Second, 
each dipole needs to have a compact geometry in such a way that its position in space can be specified with good precision, 
allowing the meaningful computation of a local density of dipoles everywhere through the system as well as the definition of 
distance dependent correlation functions. The subdivision of the system charges into dipoles cannot, but also does not need to
be permanent. In fact, the identity of the ions forming any given dipole will evolve with the changing configuration of ions in 
the ionic liquid. However, to lend credibility to the physical relevance of these dipoles, their lifetime should be long. 
Moreover, the probability distribution for the modulus $R_d$ over the dipoles population should be relatively narrow, and 
correspond to a charge separation comparable to the average distance of ions as given by a ion-ion pair distribution function.

With these considerations in mind, we set out to partition the moving charges of the system into a population of dipoles. The 
dipole moment of a set of charges is unambiguously defined provided the aggregation of charge is overall neutral. Therefore, we 
focus our attention on the dipole of neutral ion pairs, consisting of the association of one cation and one anion. The 
difficulty with this approach is that the way to partition an assembly of ions into a collection of dipoles by joining a 
[bmim]$^+$ cation and a neighbouring [NTf$_2$]$^-$ anion is not unique. The most intuitive approach could be to join each ion 
with its nearest neighbour of the opposite sign. However, the property of ion A$^+$ being the nearest neighbour of B$^-$ is not 
symmetric, since the nearest cation of B$^-$ might be C$^+$, different from A$^+$. Simple algorithms like starting from
a given ion, joining it to the closest counter-ion not already engaged into a dipole do not solve the problem. At first, this 
approach provides a series of tightly bound dipoles, satisfying the requirements listed above. Soon, however, going down the 
series of cations, the inherent frustration in the process of pairing neighbouring ions becomes manifest since the algorithm 
increasingly joins pairs separated by long distances, making meaningless the location of the dipole in space. Moreover, the
result would depend on the ordering in which the ions are paired, since the first pairs to be formed are systematically
favoured (i.e., have shorter charge separation) with respect to the last.

To correct this ambiguity, one has to define the cation-anion pairing in some physically motivated way, that makes the choice
unique and close to the intuitive idea of the fluid made of nearly equivalent neutral ion pairs of relatively short separation, 
which are easily located in space.

The approach that is proposed and tested here is to define such a unique partition of the ion population into dipoles by 
requesting that the total self-energy of all dipoles is a minimum among all possible subdivision of all cations and anions into 
neutral pairs. 

\begin{figure*}[!htb]
\begin{minipage}[c]{\textwidth}
\vskip 0.7truecm
\begin{center}
\includegraphics[scale=0.60,angle=-0]{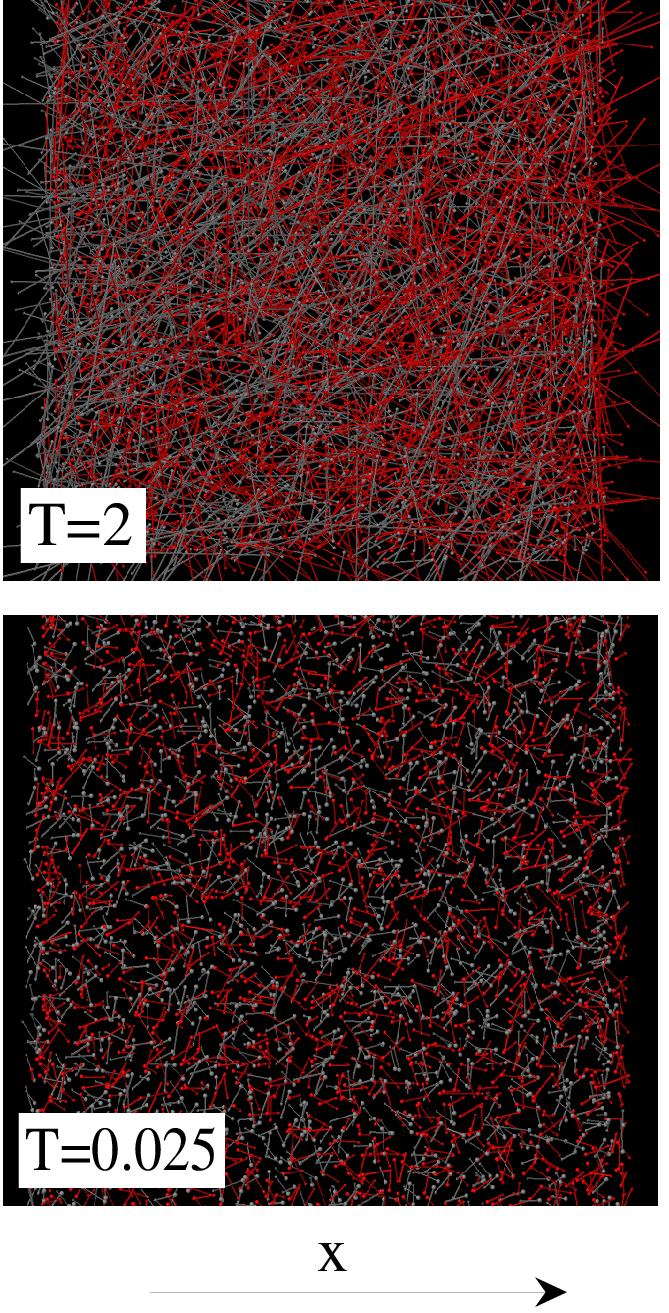}
\vskip 1.0truecm
\caption{Snapshots of the dipole configurations at the two extreme temperatures of the simulated annealing (see text) of the dry
 capacitor at $\sigma=0$. The $x$ axis is orthogonal to the two IL - graphite interfaces, with graphite not shown in the picture. Dark gray arrows point to the 
left, red arrows point to the right. $T$ is a fictitious temperature introduced to drive the MC optimisation of the ions' 
pairing into dipoles, not to be confused with the physical temperature (T=300 K) of the graphite - liquid [bmim][NTf$_2$] 
simulation. The units of the fictitious temperature are discussed in the text. 
}
\label{CapacT300}
\end{center}
\end{minipage}
\end{figure*}

Let us first consider the case of neutral graphite surfaces. The case of charged interfaces will require some slight 
modifications. The self-energy of a dipole made of charges $(q,-q)$ separated by a distance $s$ usually is given as 
$u_{s}=-q^2/s$. In the present case, in which cations and anions consists of several atoms joined into a complex ion, the 
self-energy of the dipole made of cation $\alpha$ and anion $\beta$ is defined as:
\begin{equation}
u_s(\alpha, \beta)=\sum_{i\in \alpha; j\in \beta} \frac{q_i q_j}{|{\bf r_i^{\alpha}} - {\bf r_j^{\beta}|}}
\end{equation}
where $i$ and $j$ label atoms (of charges $q_i$ and $q_j$, respectively) belonging to cation $\alpha$ and anion $\beta$, 
respectively. A very similar (although not identical) value of the self-energy is obtained by approximating $u_s(\alpha, \beta)$
with:
\begin{equation}
\tilde{u}_s(\alpha, \beta)=-\frac{1}{\mid {\bf R}^{(Q)}_{\alpha}-{\bf R}^{(Q)}_{\beta}\mid}
\end{equation}
where ${\bf R^{(Q)}}_{\alpha}$, ${\bf R^{(Q)}}_{\beta}$ are defined in Eq.~\ref{cg}. The good correspondence between 
$u_s(\alpha, \beta)$ and $\tilde{u}_s(\alpha, \beta)$ confirms the ability of Eq.~\ref{cg} to provide a meaningful coarse 
grained description of the charge distribution in the [bmim][NTf$_2$] slab, and could ease the analysis of configurations
selected from the simulation of large capacitor.

On the basis of these considerations and observations, a procedure has been devised to carry out the unbiased subdivision of 
any given configuration of ions into a configuration of dipoles, consisting of:
\begin{itemize}
\item start from the pairing of each cation with its nearest anion which is not already engaged in another pair;
\item minimize the total self-energy of the assembly of dipoles by a Monte Carlo procedures, in which neighbouring pairs of
neutral ion pairs (i.e., two cations and two anions) interchange their bonding. In other terms, given the two pairs $(i_1, j_1)$
and $(i_2, j_2)$, (fictitious) dipole bonds are swapped into $(i_1, j_2)$ and $(i_1, j_2)$ according to the energy variation:
\begin{equation}
\delta e=u_s(i_1, j_2)+u_s(i_2, j_1)-u_s(i_1, j_1)-u_s(i_2, j_2)
\end{equation}
\end{itemize}
Acceptance - rejection of the bond interchange is decided by the usual Metropolis rule at fictitious temperature $T$,
whose progressive reduction (simulated annealing) drives the system towards the minimum. The scale of he temperature interval 
and the schedule of the annealing have been decided by trial and error using a short sequence of preliminary computations.
For the sake of simplicity, electrostatic and energy units will be expresses in scaled units, in which the unit of length is
the nm, the unit of charge is $e$ and the unit of energy is $e^2/(1 nm)$, corresponding to $0.0529$ Ha, or 1.44 eV, or $139$ 
kJ/mol. In these units, a unit dipole is $48$ Debye, a unit electric field is the one producing a unit energy change on a single
positive $e$ charge upon a 1 nm displacement, etc. Also, measuring temperature in energy units, $T$ has been decreased from 
$T=2$ to $T=0.025$ energy units in 80 runs of $64\ 10^6$ MC steps, each consisting of the attempted 
interchange of bonds between neighbouring dipoles. Since the attempted MC move is a discrete event that cannot be fractioned,
the acceptance ratio of the MC cannot be tuned, but it exceeded $0.5$ at high $T$, and decreased to $5\ 10^{-5}$  (one accepted
move every 20 thousand moves) at the lowest temperature, i.e., $T=0.025$.

\begin{figure*}[!htb]
\begin{minipage}[c]{\textwidth}
\vskip 0.7truecm
\begin{center}
\includegraphics[scale=0.60,angle=-0]{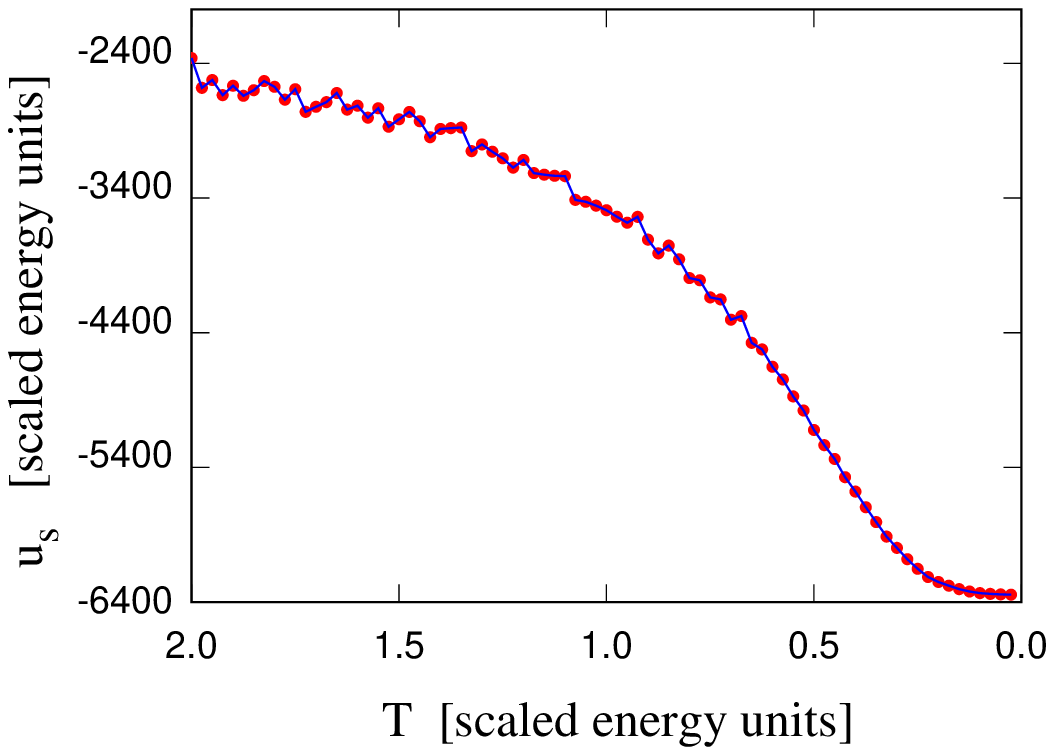}
\vskip 1.0truecm
\caption{Dipoles' self-energy $u_s$ as a function of $T$ during the annealing. Uncharged capacitor at $T=300$ K. The scaled 
units for $u_s$ and $T$ are defined in the text. The $u_s$ values refer to the self energy of all $3072$ ion pairs.
}
\label{udip}
\end{center}
\end{minipage}
\end{figure*}

The effect of the MC annealing on the definition of dipoles on top of a single configuration of the ions is illustrated
in Fig.~\ref{CapacT300}. Arrow represent the strength and spatial location of each dipole in the sample. The choice of the 
location of each dipole is explained in the following paragraph. The bottom panel shows the final configuration resulting from 
the simulated annealing. The configuration used as an example has been extracted from the simulation of the capacitor
at $T=300$ K, $P=1$ bar, $\sigma=0$. The topmost panel shown a configuration of dipoles selected at random during the simulation
at the highest temperature ($T=2$) of the simulated annealing protocol. The color of each arrow is assigned as follows: if the 
$x$ component of the dipole points to the right, the arrow is painted red, otherwise it is painted gray. The number of red
dipoles equals the number of gray dipoles to within $1$ \%. From the figure, it is apparent that at the end of the MC procedure,
dipoles are of similar size and the sum of their self-energies, although unlikely to be the true absolute minimum, is already 
low in energy (see the self-energy as a function of $T$ during the annealing in Fig.~\ref{udip}). It might be worth emphasising 
that this procedure is carried out on a single configuration of the sample, i.e., at fixed position of all atoms in the system, 
only the distribution of fictitious dipole bonds among ions being interchanged.

\begin{figure*}[!htb]
\begin{minipage}[c]{\textwidth}
\vskip 0.7truecm
\begin{center}
\includegraphics[scale=0.60,angle=-0]{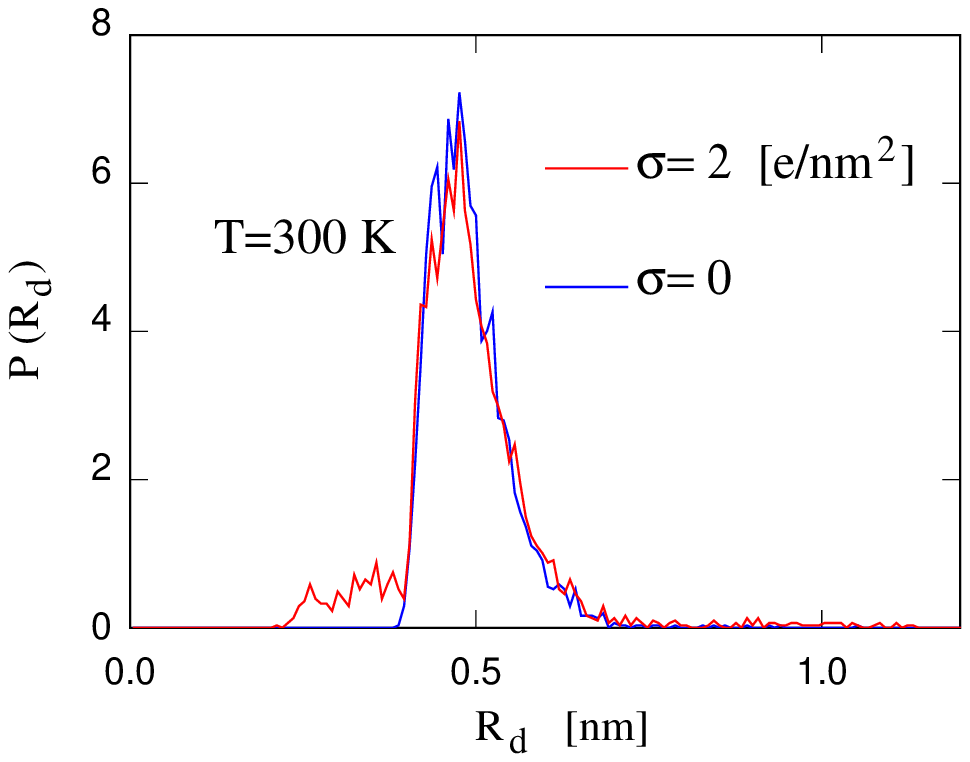}
\vskip 1.0truecm
\caption{Probability distribution for the charge separation of the dipoles at $T=0.025$ (end of the annealing) for
the neutral interface (blue line) and for the charged interface at $\sigma=2$. The two distributions have been
computed from the two dipole populations shown in Fig.~\ref{CapacT300}. The area under the curve is normalised to:
100 in the case of the neutral interface; $100 \times (3072+239)/3072$ for the charged interface, to account for the
239 ion pairs added to represent the charge on the electrodes.
}
\label{prob-dip}
\end{center}
\end{minipage}
\end{figure*}

The probability distribution for the separation $R_d$ of the charges for the dipoles in the final configuration is shown in 
Fig.~\ref{prob-dip}. The average value $\langle R_d \rangle=0.5$ nm corresponds  well to the geometric separation of cations and
anions in the radial distribution function of the homogeneous [bmim][NTf$_2$] liquid at the same conditions. The full width at 
half peak of about $0.1$ nm is sufficiently short to support the idea that the dipoles defined in this way, although not 
permanent, are indeed physical associations of ions and not only fleeting geometrical details in the evolving ionic 
configuration of the [bmim][NTf$_2$] sample.

\begin{figure}[!htb]
%\begin{minipage}[c]{\textwidth}
%\vskip 0.7truecm
%\begin{center}
\includegraphics[scale=0.70,angle=-0]{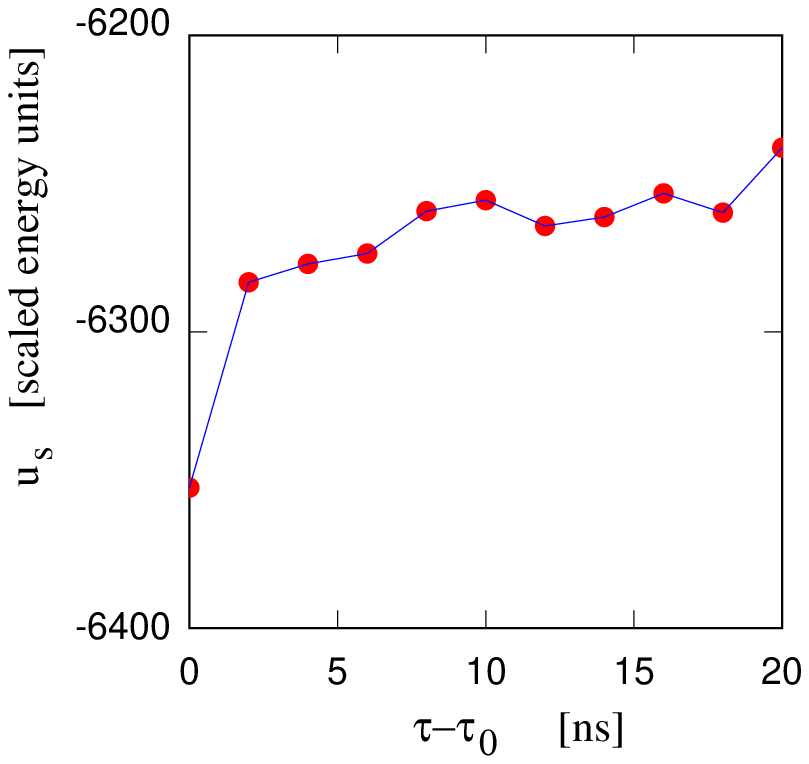}
%\vskip 1.0truecm
\caption{Self-energy $u_s$ of the optimal ion pairing determined for the MD configuration at time $\tau_0$, ri-evaluated 
on the MD configuration at time $\tau>\tau_0$. In other terms, a list of ions pairs minimising the self-energy $u_s$ is
determined at time $\tau_0$. Then, the self-energy expression for the same ion pairs is ri-evaluated using
the ions' coordinates extracted from the MD trajectory at time $\tau$. The $u_s$ value refers to the total population of
$3072$ ion pairs.
}
\label{ene0}
%\end{center}
%\end{minipage}
\end{figure}

It can be supposed that a variety of subdivisions into pairs have comparable and relatively low self-energy, all representing
the properties of the ideal unique and best pairing choice. However, repeated pairing optimisations, carried out by starting 
from different randomized starting points had a minimum fraction of equal pairs at the end of the simulated annealing of 66 \%,
showing that the self-energy basin around the (unknown) ideal minimum is relatively wide, but still well defined, without
a plurality of significantly different local minima making too ambiguous the pairing of ions into neutral pairs.

\begin{figure*}[!htb]
\begin{minipage}[c]{\textwidth}
\vskip 0.7truecm
\begin{center}
\includegraphics[scale=0.60,angle=-0]{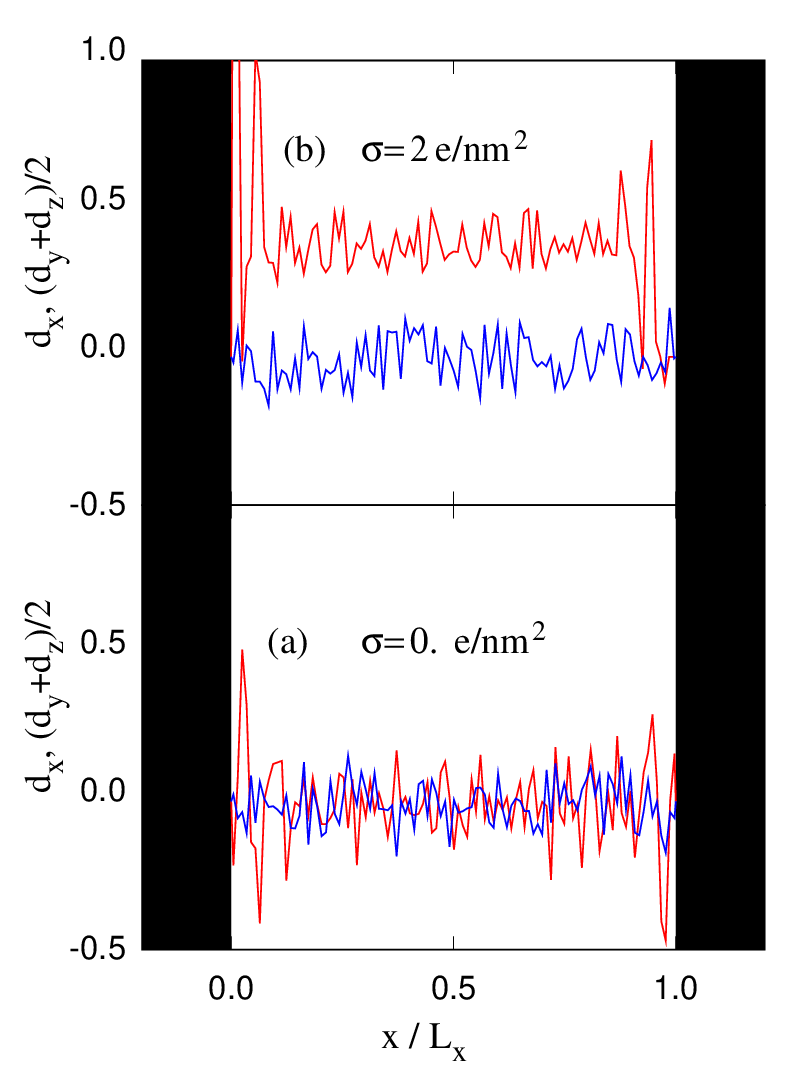}
\vskip 1.0truecm
\caption{Density of dipole as a function of position along the $x$ axis. The red lines show the $\langle d_x\rangle$ component, 
while the blue line shows the average $\langle d_y+d_z\rangle / 2 $ of the $d_y$ and $d_z$ components. The black rectangles 
represent the graphite electrodes. When $\sigma \neq 0$, the average of $\langle d_x\rangle$ does not vanish far from the
interfaces.
}
\label{ax}
\end{center}
\end{minipage}
\end{figure*}

Since the grouping of ions into dimers is the result of a probabilistic algorithm that runs independently on different 
configurations, it is difficult to estimate an accurate value for the dipoles' lifetime. However, it has been verified that 
the ion pairing resulting from optimisation protocol on the MD configuration at time $\tau_0$ remains a low self-energy pairing
for ions' configurations extracted at later times $\tau > \tau_0$ for $\tau -\tau_0$ of the order of tens of ns, as can be
seen in Fig.~\ref{ene0}. This observation suggests a lifetime of a similar duration for the dipoles. 

The compact geometry of the ion pair carrying the dipole (say, dipole $\alpha$) allows us to define its geometric location 
through the relation:
\begin{equation}
r^{dipole}_{\alpha}=\frac{\sum_{i\in \alpha} \mid q_i\mid {\bf r_i}}{\sum_{i\in \alpha} \mid q_i\mid}
\end{equation}
This allows to compute the density of dipole moment along each direction, which is displayed in Fig.~\ref{ax} (a) for the sample
illustrated in Fig.~\ref{CapacT300}, having $\sigma=0$ surface charge density on the graphite electrodes. It is apparent that 
the distribution of dipole moments vanishes over most of the sample, with possibly narrow peaks in  proximity of the two
interfaces.

The visual inspection of the dipole configuration in Fig.~\ref{CapacT300} does not show any clear regularity. A more
quantitative analysis involving the determination of the structure factor for the mixture of red and gray dipoles might still
reveal the presence of clusters of dipoles pointing towards the same interface, providing at the same time an estimate of the
size of these hypothetical domains.

The same protocol could be used to pairs ions into dipoles when the capacitor plates carry a non vanishing surface charge
density. The resulting configuration of dipoles, however, is less regular than the one obtained at $\sigma=0$, and the MC 
optimisation is less able to provide a low energy partition of ions in dipoles of nearly homogeneous length. The reason is that
the excess of ions of a given $\pm e$ charge screening the $\mp \sigma$ charge on the plate cannot find a neighbouring partner in 
their immediate vicinity, being forced to form dipoles with ions located far into the IL bulk, or even close to the opposite
plate, there an excess of counter ions can be found. In fact, at the end of virtually every simulated annealing cycles, the 
dipoles population included a few long separation (high self-energy) dipoles, unable to escape a local minimum of the total 
self-energy, and only a sustained effort is able to cure these configurations through very long simulated annealing cycles.

To overcome this problem, the original ions' population is supplemented by further ions, whose role is to represent the charge 
on the capacitor plates, providing a pairing opportunity for the counter-ions that populate the first IL layer next to the 
interface. This procedure is better illustrated with an example. Let us consider the same IL sample of $3072$ ion pairs, 
enclosed in between two graphite electrodes whose interfacial layers, each consisting of $4784$ carbon atoms, now carry 
$\pm 0.05\ e$ per C atom, for a total of $\pm 239.2$ e on the two capacitor plates. In the simulation, the left plate is 
negatively charged, and the right plate is positively charged. Since each carbon atoms carries only $\pm 0.05$ e, the 
representation of the charge on each electrode cannot be obtained by including all these atoms in the dipole formation protocol.
Instead, a number of integer charges corresponding to the total charge on each electrode is introduced. In the present example 
$239$ $(-e)$ charges are located on the left electrode, and 239 $(+e)$ charges are positioned on the right electrode. 

At the beginning, cations are sorted in order of increasing distance from the anode. Then, each of the $239$ cations closest to
the anode, of coordinates $\{ {\bf r_i}, i=1, ..., 239\}$ is paired with a $(-e)$ charge located $0.2$ nm below the graphite 
plane, at the same $(y_i, z_i)$ in-plane position of the corresponding cation. The specular operation is carried out with
the opposite polarity electrode, interchanging the role of cations and anions, using the distance from the positively charged
plate to select the anions to be paired with the surface positive charge.

After this first $239+239$ ion pairs are formed, linking ions of opposite sign across the interface, the remaining 2594 ion 
pairs are given an initial pairing into dipoles precisely as in the $\sigma=0$ case. Then, the MC annealing is carried out
following the same schedule as in the neutral case. Bond interchanges concern both the original ions, and the ions added to
represent the surface charge.

Following this approach, the optimization of the self-energy runs as smoothly and as effectively as in the neutral case, 
producing a probability distribution of dipole separations $R_d$ virtually superimposable to the one from the neutral case. The 
only noticeable difference is a tail
at short separations apparent in the $\sigma=2$ e/nm$^2$ plot, due to the tight pairs joining the added surface charges to the
counter-ions in the inner screening layer (see Fig.~\ref{prob-dip}). In the annealed configuration, 85 \% of the dipoles point 
towards the right electrode. As apparent from the figure, the dipole density has a net excess of $d_x$ dipole moment along the 
entire capacitor along $x$. The net sum of all dipoles along $x$ has a direction which is such to partly balance the dipole 
formed by the two graphite surfaces of $\pm \sigma$ charge density.
The observation of an excess of dipole moment far from charged interfaces is the major result of the present investigation.

Also in this case, repeating the simulated annealing after randomising the pairing of ions results in a set of ion pairs 
whose overlap with any previous one is very sizeable. In fact, the overlap seems to be even stronger than in the neutral case,
perhaps suggesting that the minimum energy basin is slightly narrower and somewhat simpler for charged interfaces than
for the neutral ones.

\begin{figure*}[!htb]
\begin{minipage}[c]{\textwidth}
\vskip 0.7truecm
\begin{center}
\includegraphics[scale=0.60,angle=-0]{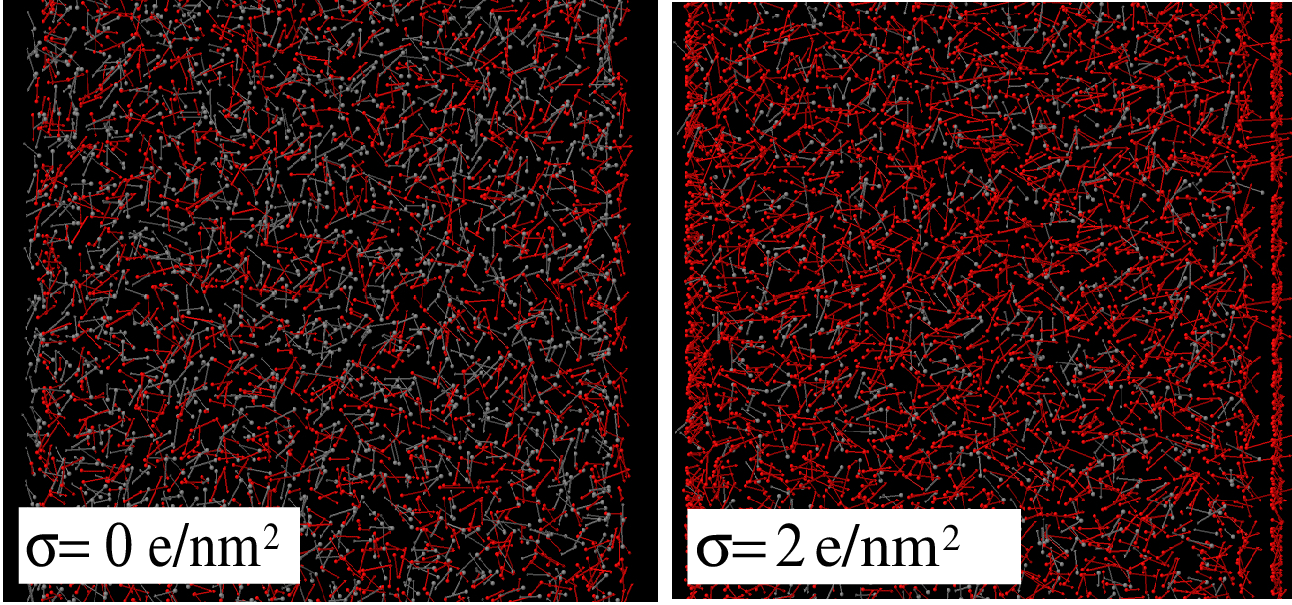}
\vskip 1.0truecm
\caption{Dipoles' configuration at he end of the simulated annealing optimization as a function of surface charge density
$\sigma$.  Red dipoles point to the right side of the figure, white dipoles point to the left. The prevailing direction of 
dipoles
in the charged interface case is such to partially compensate the dipole moment of the two charged graphite layers
forming the plates of the capacitor.
}
\label{sigma}
\end{center}
\end{minipage}
\end{figure*}

A quantitative assessment of the dipoles' alignment along $\hat{x}$ is given in Fig.~\ref{ax}, showing the average dipole moment
density
along the directions parallel and normal to the interface. The data reported in this figure refer to the dipole configurations
shown in Fig.~\ref{sigma}. Panel (a) in Fig.~\ref{ax}, in particular, refers to the $\sigma=0$ sample in Fig.~\ref{sigma}, while
Panel (b) in Fig.~\ref{ax} refers to the $\sigma=2$ $e/nm^{2}$ sample in the same Fig.~\ref{sigma}. The $\langle d_y \rangle$
and $\langle d_z \rangle$ components of the dipoles, on average, vanish across the whole capacitor in both samples. Again in 
both samples, the $\langle d_x\rangle$ component is affected by the presence of the interface over a short contact layer of
sub-nanometric width. The important novelty is that $\langle d_x\rangle$ does not vanish well inside the [bmim][NTf$_2$] layer 
in the $\sigma=2$ $e/nm^{2}$ case.

\subsection{Speculations on the mechanism connecting the dipoles' configuration and the spectroscopic properties}
\label{speculation}

\begin{figure*}[!htb]
\begin{minipage}[c]{\textwidth}
\vskip 0.7truecm
\begin{center}
\includegraphics[scale=0.60,angle=-0]{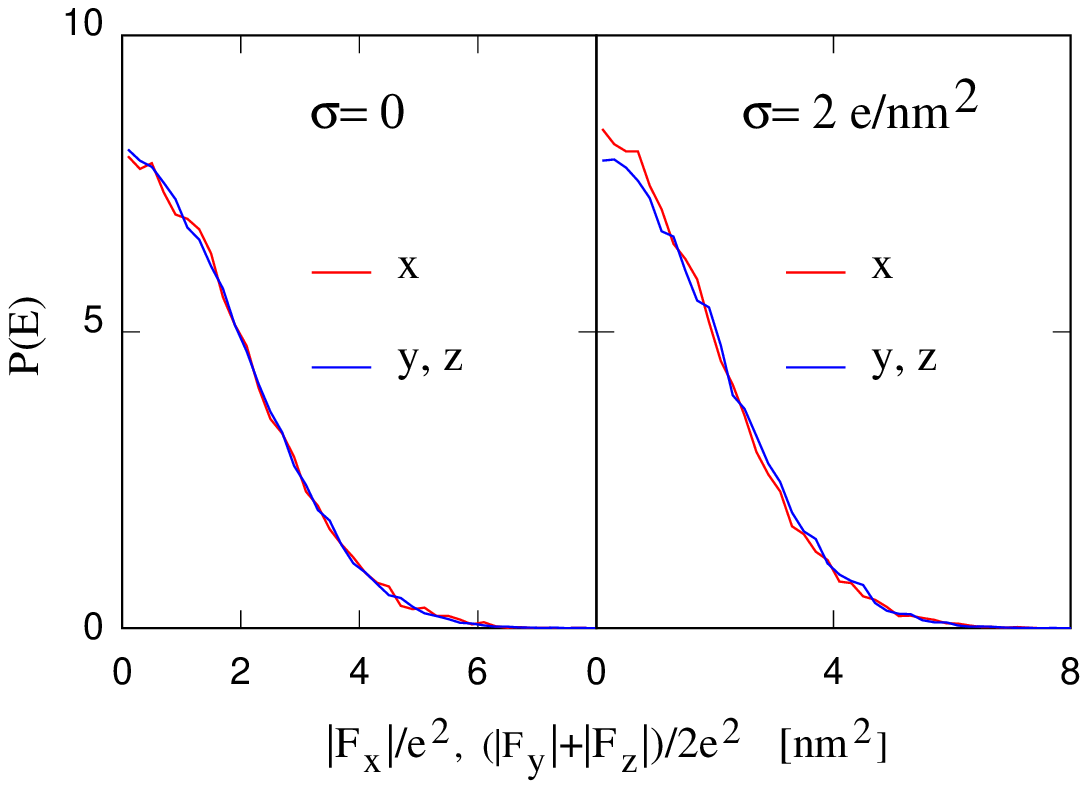}
\vskip 1.0truecm
\caption{Probability distribution for the Cartesian components of the force ${\bf F_{\alpha}}$ acting on each ion pair $\alpha$
in the cental portion of the [bmim][NTf$_2$] slab (see text) due to all other ion pairs in the sample. Comparison of the results
for the capacitor at: $\sigma=0$ and $\sigma=2$ $e/nm^{2}$ surface charge density.
}
\label{efluct}
\end{center}
\end{minipage}
\end{figure*}

The discovery of a geometric parameter such as $\langle d_x\rangle$ whose non-vanishing value carries information on the
presence of a charged interface well inside the [bmim][NTf$_2$] slab represents a novel and interesting observation, which goes 
beyond the established description of electrified interfaces. In itself, however, this observation does not explain the changes 
in spectroscopic properties revealed by a few experiments. In fact, the computational and experimental results might have no 
obvious connection with each other apart from probing the same system. Nevertheless, it is conceivable that the parallel
orientation of instantaneous dipoles due to a charge on the capacitor could reveal a deeper aspect which could explain both the 
dipoles ordering and the change in optical properties. This linking aspect needs to be compatible with the classical results of 
static screening, but otherwise could affect the ionic liquid properties as far as the correlation of the instantaneous dipoles 
could penetrate. A conceivable scenario concerns the properties of the fluctuating electric fields acting on the ions and 
evolving according to their thermal motion. Far from the graphite electrodes, the average electric field has to vanish, since 
[bmim][NTf$_2$] is a conductor, but no obvious constraint limits the size as well as the position and size variations of 
fluctuating electric fields.

To verify this scenario, we identified all dipolar IL ion pairs whose cation is (instantaneously) located within the central $5$
nm of the capacitor. For each pair, the total electrostatic force acting on the cation and corresponding anion was computed,
accounting for all other ions in the sample, i.e., not only those in the central $5$ nm slab. The surface charge density on the
electrodes is not included in this computation, whose target is the quantification of the fluctuating electric fields in the 
bulk region of the ionic liquid. Since the simulated samples are large, the electrostatic force on the ions accounted for the 
complementary error part of the Ewald energy only, neglecting all reciprocal space terms. This approximation is introduced for 
the sake of simplicity, but its accuracy is enhanced by selecting the decaying length of the complementary error function in 
such a way to prevent the interaction of the central ions with the interfaces, but otherwise to be as large as possible to 
minimize the size of reciprocal space terms.

The computation of the electric fields has been carried out on a population of independent configurations for the samples at 
$\sigma=0$ and $\sigma=2$ $e/nm^2$. The results are illustrated in Fig.~\ref{efluct}. They can be summarised as follows. In both
cases, the average force vanishes to the same accuracy in the three Cartesian directions, confirming that the screening of 
external electric fields is verified to within the accuracy of the computation. The properties of electric field fluctuating in 
time and space are characterised by computing the distribution probability for the intensity of the three Cartesian components 
of the force on each dipole. Both for the $\sigma=0$ and the $\sigma=2$ $e/nm^2$ capacitors, the probability distribution for 
the forces along $\hat{y}$ and $\hat{z}$ are the same, again to within the computational accuracy, and the values of the two
probability distributions are averaged together to improve the statistics in the comparison with the result in the $\hat{x}$ 
directions. As apparent in Fig.~\ref{efluct}, the probability distribution for $| F_x |$ cannot be distinguished from the
$(|F_y|+|F_z|)/2$ case for the uncharged capacitor. A slight difference appears in the charged capacitor case, localised in the
range of low fields. The height of the two peaks differ by about $8$\%, which is only slightly less than their combined error 
bar. As a consequence of the slight difference in $P(F)$, the average value of $| F_x |$ is about $1.5$ \% lower than the 
average value of $(|F_y|+|F_z|)/2$, again comparable to the combined error bar on the two quantities that are being compared. 
Increasing statistics to decrease the statistical error bar well beyond the separation of these quantities would require 
carrying out simulations and configuration analyses at least for times longer than those already done, a target that would 
require a dedicated effort as well as dedicated computational resources. To be safe, we can state that, at this stage, the 
computation is not conclusive, neither definitely supporting and certainly not disproving that the geometric properties of the 
dipoles population and the spectroscopic properties are related by changes in the fluctuating electric field inside the ionic 
liquid. Nevertheless, the scenario outlined in this section remains appealing, since it does not violate any known condition and
it is not disproved by the present simulations. 

\section{Summary and discussion}
\label{summary}

The interface between ionic liquids and solid electrodes is a fascinating subject that joins the current interest for a vast 
number of electrochemical applications with concepts and theories dating back to one century ago.\cite{korn1, korn2, korn3, 
bock}. This fertile combination, together with the complex structure of ions and the nanostructured texture of ionic liquids, 
has fostered a variety of results of practical interest, and also offered genuine surprises that challenge the accepted 
picture of the electrified interfaces. Among the aspects that are being challenged is the long held notion that, beyond a 
narrow interfacial region of microscopic width, a Coulombic fluid in contact with a charged interface is virtually bulk-like. 
A strong indication in this direction is provided by the recent observation\cite{expLangmuir} of the Pockel effect at the 
interface between the [bmim][NTf$_2$] ionic liquid and two planar solid parallel electrodes connected to a generator maintaining
an electrostatic potential energy difference between the two sides of the IL slab. Upon charging the interface, a change in the 
Raman intensity was observed, arising from all Raman-active modes of cations and anions, independently from their distance 
from the interface.

To investigate this phenomenon, the interface between the same [bmim][NTf$_2$] ionic liquid and a graphite electrode has been
simulated by molecular dynamics, using an atomistic semi-empirical force field. An innovative analysis of trajectories
allows to join [bmim]$^+$ and [NTf$_2$]$^-$ ions into neutral pairs, whose dipoles show an apparent alignment with the external
electric field driven by the charge deposited on the graphite surface. This alignment is observed even well inside the IL 
bulk-like region, where the external field is exactly cancelled (on average) by the screening at the interface by the mobile
ions in the IL. Since dipoles do not obey the same strict screening conditions of charges, the observation does not violate any 
accepted theory of electrified interfaces, but opens the way to a whole new class of medium-long range correlations that remain 
hidden when one considers the system only as a collection of free ions.

Up to now, this observation concerns a geometric property in the distribution of ions throughout the system, that leaves open 
questions concerning its origin and its significance with respect to the observed effect on the Raman spectra measured on
the experimental samples. The main question on its origin concerns the role of the shape anisotropy of the ions in transferring
the information on the state of the interface deep into the IL bulk. The remaining question on the relevance of the geometric 
property concerns whether this correlation among virtual dipoles can alter the vibrational properties of the ionic liquid. The 
mechanism joining the geometric correlation with the change of Raman spectroscopic properties has to respect the fact that the 
average 
electric field well inside the ionic liquid vanishes, independently from the charge on the electrode. This fact, however, does 
not prevent changes in the average {\it square amplitude} of the fluctuating electric field inside the ionic liquid, that could 
indeed give origin to the observed change of vibrational frequencies. A preliminary investigation of this scenario by MD did
not provide a definite answer, since the effect is likely to be small and acquiring sufficient statistics requires a very
intensive effort.

Because of these open questions that are currently being investigated, the present document represents a report on work in
progress, that will be followed by a full manuscript on the comprehensive investigation of the [bmim][NTf$_2$] / graphite
interface and on the hidden correlations in the distribution of ions arising upon charging the electrodes.

Further considerations that will be explored during later stages of our investigations are as follows. Long range correlations 
imply that the two electrodes in simulated planar capacitors are likely to interact up to the largest sizes that are practically
achievable right now by computer simulation, affecting any estimate of properties that depend on the spatial distribution and orientation of the ions. 
This problem enhances the appeal of single-electrode schemes like the one introduced in Ref.~\onlinecite{eletransf}, in which
a unique spherical electrode surrounds the pool of electrolyte, whose thermodynamic state is controlled by adopting the grand
canonical simulation approach.

The observation of a peculiar geometric property that conveys information on polarisation from the interface to deep into
the bulk electrolyte is reminiscent of the role of the Berry phase in ab-initio investigations of polarisation in
ferroelectric crystals.\cite{raffaele} Although the Berry phase is, to our knowledge, an exclusively quantum mechanical 
feature, it might be worth exploring the analogy with the long range correlations highlighted in the present study.
 
As a last remark, we observe that our geometric construction building (instantaneous) dipoles out of a population of mobile ions
has some points of contact with the derivation in Ref.~\onlinecite{bazant} of a spin-glass Hamiltonian mimicking structural
properties of viscous ionic liquids, which is then used to characterize a variety of correlation properties. The difference is 
that, while Ref.~\onlinecite{bazant} emphasizes many-body concepts, the present approach closely follows traditional lines of
electrostatics, defining new entities (dipoles) from simpler components like ions. This last procedure makes sense in this case,
in which ions are complex, their motion is sluggish at best, and combining them into dipoles decreases the range of their 
interaction, opening the way to the types of correlations discussed in the present paper.

ACKNOWLEDGMENTS: N.C.F.-M. gratefully acknowledges the ECHELON Project: P2019-02-003 from the Carl Zeiss Foundation for financial support. D.V.-D., and N.C.F.-M. acknowledge the Deutsche Forschungsgemeinschaft (DFG, German Research Foundation) for funding through the research group FOR 2982–UNODE, Project number 413163866. R.C.-H. acknowledges funding from SFB-TRR146 of the German Research Foundation (DFG)–Project No. 233630050 and the European Union funding through the Twinning project FORGREENSOFT (grant no. 101078989 under HORIZON-WIDERA-2021-ACCESS-03). D.V.-D., R.C.-H., and N.C.F.-M. thank Kurt Kremer and Friederike Schmid for useful discussions.

\end{document}